\documentclass{iau}

\usepackage{amsmath}
\usepackage{graphicx}
\usepackage{multirow}

\begin{document}

\lefttitle{J. Jurysek et al.}
\righttitle{SST-1M: Recent results and prospects for observation of the Galactic Center region}

\jnlPage{1}{7}
\jnlDoiYr{2026}
\doival{10.1017/xxxxx}

\aopheadtitle{Proceedings IAU Symposium}
\editors{M. Zaja\v{c}ek,  T. Je\v{r}\'{a}bkov\'{a}, V. Karas, R. Schödel \&  P. Sukov\'{a}, eds.}

\title{SST-1M Cherenkov telescopes: \\Recent results and prospects for observation of the Galactic Center region}

\author{
J. Jury\v{s}ek on behalf of the SST-1M Collaboration\hyperref[authorlist]{*}
}
\affiliation{
FZU - Institute of Physics of the Czech Academy of Sciences, Na Slovance 1999/2, Prague 8, Czech Republic
}

\begin{abstract}
The Galactic Center is a crowded region containing powerful particle accelerators and objects producing non-thermal radiation, including the diffuse very-high-energy (VHE) gamma-ray component known as the `Ridge', whose hard spectrum suggests particle acceleration up to PeV energies. The relevant processes can be tested, and the source parameters can be constrained via Cherenkov telescope observations.
Two Single-Mirror Small-Size Cherenkov Telescopes (SST-1M) are currently operated in stereoscopic mode at the Ondřejov Observatory. Their future relocation is being considered, including a site with excellent visibility of the Galactic Center. We present recent observations demonstrating SST-1M capabilities for detecting extended emission and discuss prospects for probing the Galactic Center region, including searches for a spectral cut-off above 10 TeV. We show that the large field-of-view and good VHE sensitivity make SST-1M an ideal instrument for the search for PeVatrons.
\end{abstract}


\maketitle

\section{Introduction}

The center of the Galaxy is a complex region, rich in dense molecular clouds, supernova remnants, and massive stellar clusters with active star formation, all of which are potential sites of particle acceleration and sources of non-thermal radiation above 100 GeV. The very-high-energy (VHE) emitters in the region include several point-like or slightly extended sources, such as HESS J1745-290 \citep{2004A&A...425L..13A, 2016Natur.531..476H} and VER J1745–290 \citep{2016ApJ...821..129A}, associated with Sgr A*, SNR G0.9+0.1 \citep{2005A&A...432L..25A}, and unidentified sources, such as VER 1746–289/HESS J1746–285 \citep{2014ApJ...790..149A, 2018A&A...612A...9H}. These sources are located within a diffuse emission component, also called the `Ridge', making the identification of the origin of the VHE gamma-ray emission challenging due to difficulty with source confusion \citep{2016Natur.531..476H, 2021ApJ...913..115A, 2020A&A...642A.190M, 2024ApJ...973L..34A}.

The spectral properties of the diffuse emission remain uncertain; nonetheless, they bear great potential to constrain the composition of the emitting region and discriminate between leptonic vs. hadronic scenarios. \citet{2016Natur.531..476H} reported a hard power-law (PL) spectrum for the diffuse component without any spectral cutoff up to tens of TeV, later confirmed by VERITAS \citep{2021ApJ...913..115A}. In the case of hadronic origin of the emission, it could indicate particle acceleration up to PeV energies. \citet{2020A&A...642A.190M}, on the other hand, reports a $2\sigma$ cutoff at $20^{+60}_{-10}$~TeV. HAWC \citep{2024ApJ...973L..34A} cannot resolve the region due to limited angular resolution, but detected photons above 100 TeV with no significant cutoff, yet a softer PL index ($\Gamma = -2.88$) compared to H.E.S.S. and VERITAS. LST-1 \citep{2025icrc.confE.542A} recently resolved the diffuse emission and claims a $3\sigma$ hint of the cutoff at $20^{+20}_{-10}$~TeV. Last but not least, H.E.S.S. Collaboration \citep[][in these proceedings, submitted 2026]{2025arXiv250700132D, hess_these_proc} presented an updated analysis of the 16-yr data sample, showing a hint ($3\sigma$) of a spectral curvature or cutoff. The situation remains puzzling, mainly due to difficult observations at large zenith angles for Northern observatories and the source confusion.
 
The Single-Mirror Small Size Telescopes (SST-1Ms) are 4-m-class Cherenkov telescopes designed to detect gamma rays at multi-TeV energies \citep{sst1m_performance_paper, sst1m_hw_paper}. Two SST-1M telescopes separated by 155.2 m are currently operated at the Ond\v{r}ejov Observatory, Czech Republic (510 m a.s.l.). Since 2023, stereoscopic observations of various astrophysical gamma-ray sources have been performed, proving the exceptional performance of the observatory for the detection and study of extended sources of gamma rays beyond 1 TeV \citep{sst1m_performance_paper, 2025arXiv250716408J}. However, the northern latitude of the site does not us to observe the Galactic Center. Future relocation of the telescopes to Malarg\"ue in Argentina is being considered to exploit the potential of the SST-1M observatory for Galactic Center science.

In this contribution, we summarize recent highlights from the SST-1M operation in Ond\v{r}ejov and discuss the prospects for Galactic Center observations when relocated to the Southern Hemisphere.

\section{Recent scientific highlights of SST-1M}

Since the beginning of stereoscopic operation in 2023, the SST-1M telescopes have collected 765 hours of stereo data, composed of benchmarking observations of the Crab Nebula, studies of Galactic PeVatron candidates visible from Ond\v{r}ejov, blazar monitoring, and transient follow-ups \citep{2025icrc.confE.746T}. 

\begin{figure}[t]
\centering
\begin{tabular}{cc}
\includegraphics[width=.45\textwidth]{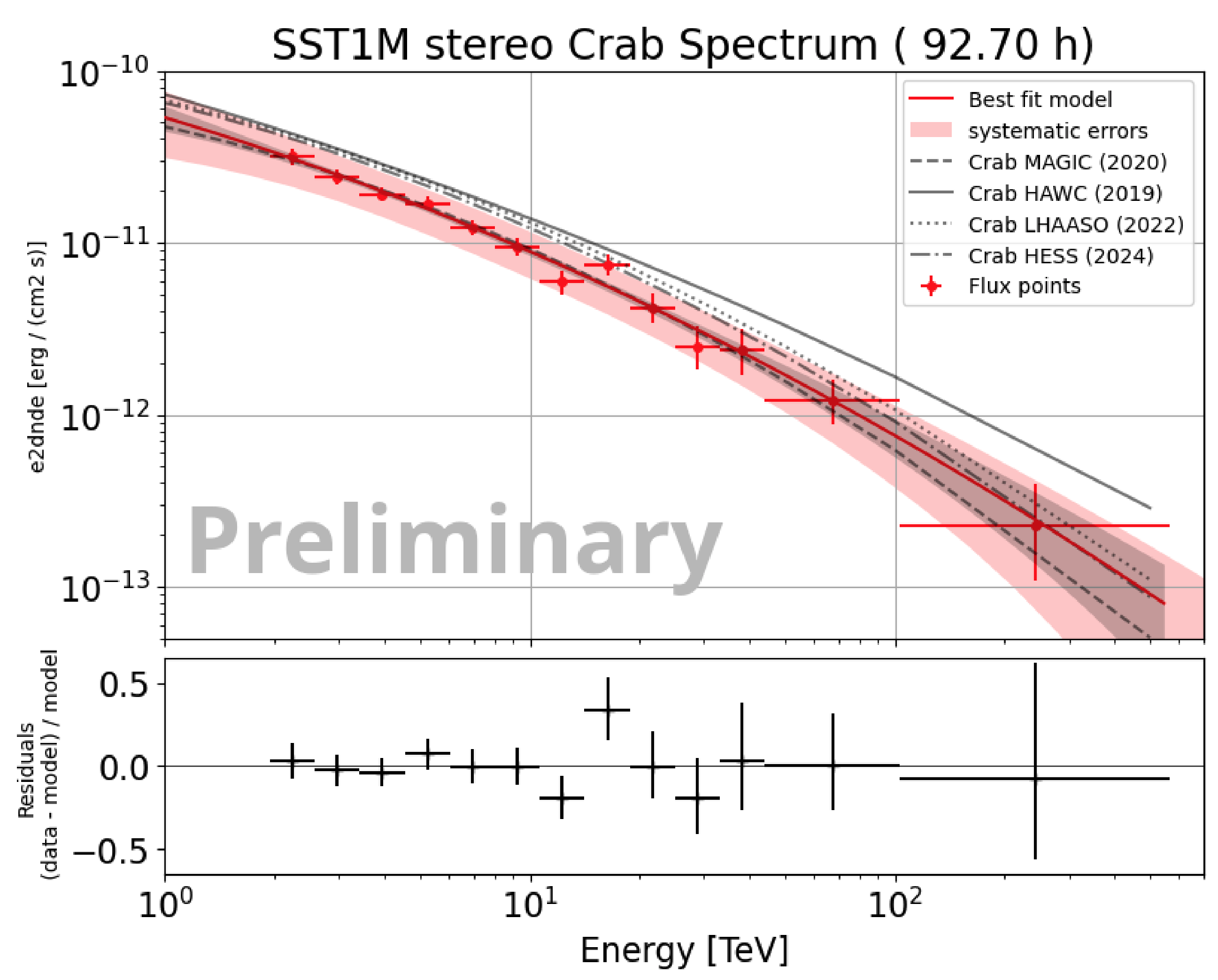} &
\raisebox{0mm}{
\includegraphics[width=.48\textwidth]{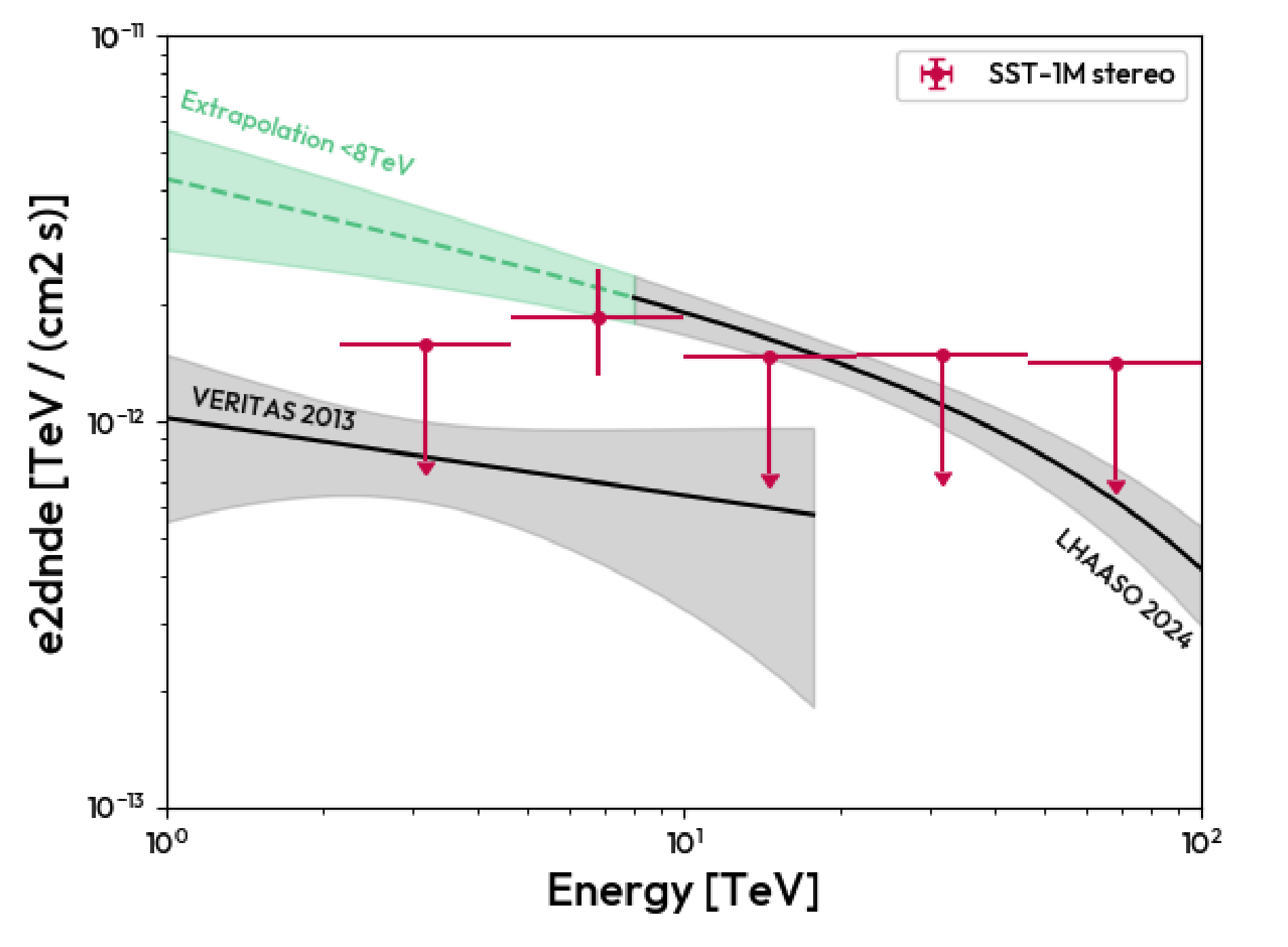}} \\

\includegraphics[width=.50\textwidth]{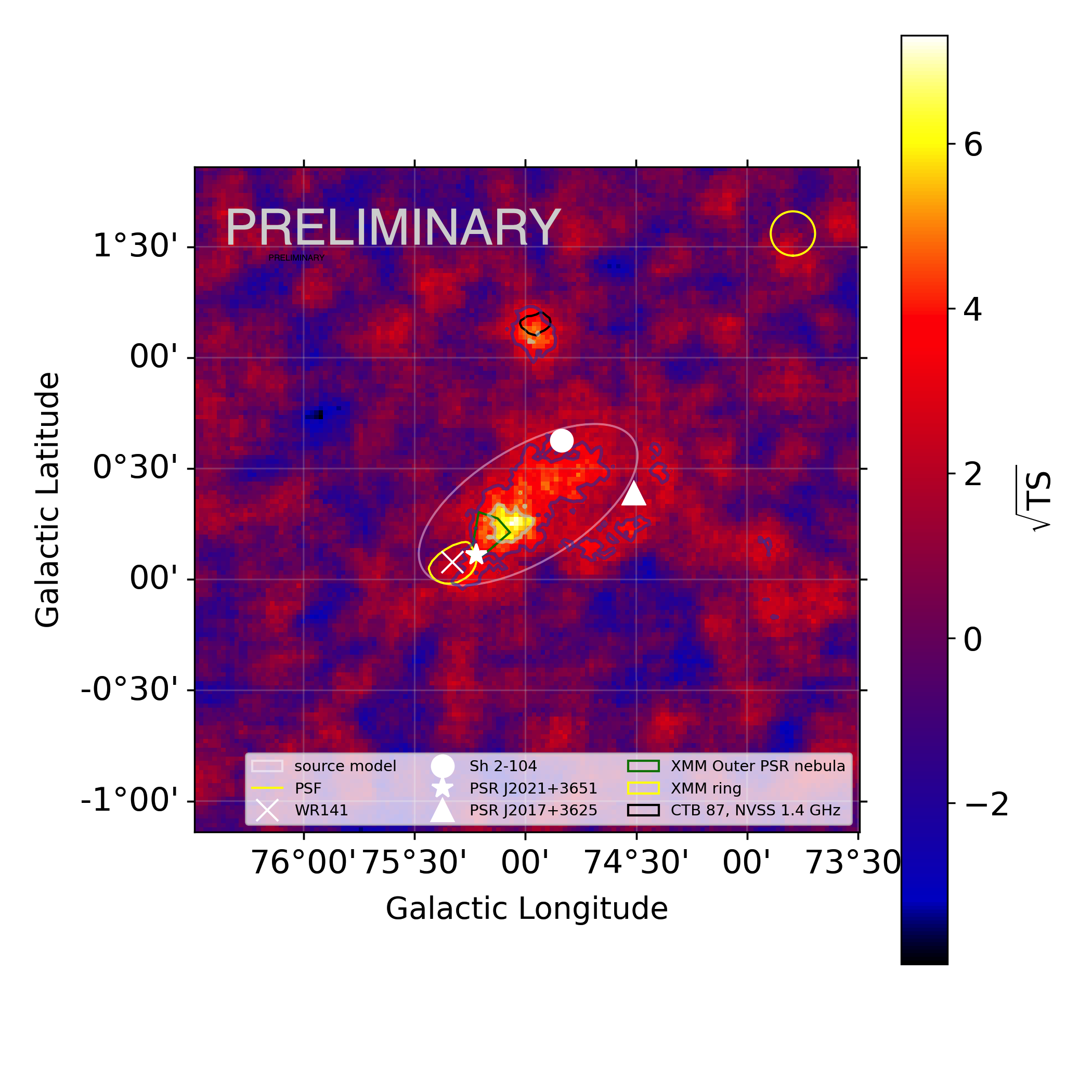} &
\raisebox{8mm}{
\includegraphics[width=.45\textwidth]{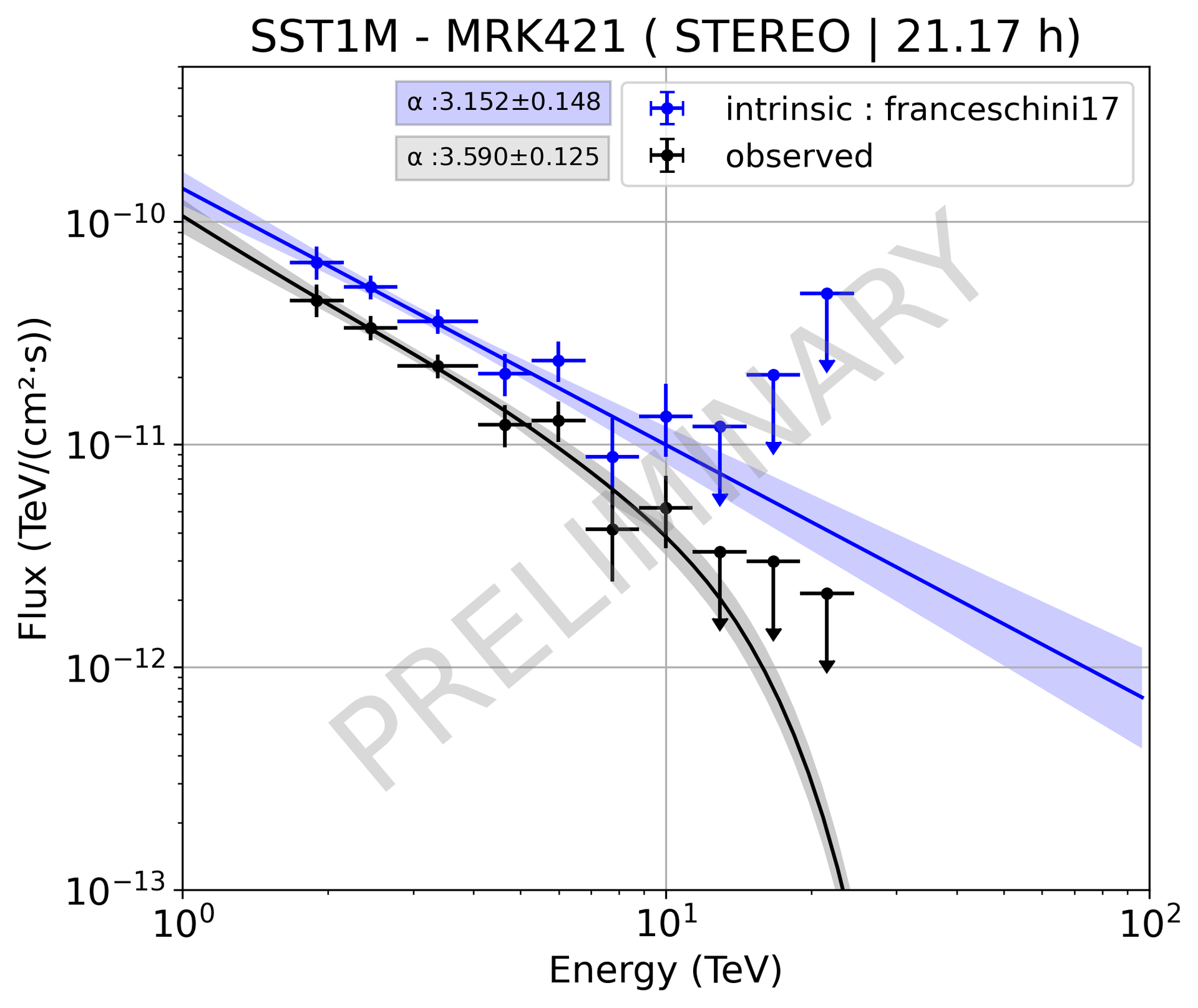}} \\
\end{tabular}
\caption{\textit{Upper Left:} Spectral energy distribution of the Crab Nebula  \citep{2025arXiv250808861T}. \textit{Upper Right:} Spectral Energy Distribution of CTA 1 compared with results of other observatories \citep{2025arXiv250926068L}. \textit{Bottom Left:} Galactic PeVatron candidate VER J2019+368 as seen with the SST-1M in 120 hours \citep{2025arXiv250716408J}. The yellow circle represents the PSF of the instrument. The white contour represents the best-fitting spatial model (68\% containment). \textit{Bottom Right:} Averaged spectrum of Mrk 421 during increased activity in January 2026.}
\label{fig.results}
\end{figure}

Crab Nebula observation for 93 hours resulted in the spectrum measured up to 100 TeV (see Fig.~\ref{fig.results}, upper left panel), with the detection significance of $45\sigma$ ($4.4\sigma$ for $E>50$~TeV) \citep{2025arXiv250808861T}. The collected dataset served to validate the Monte Carlo (MC) model of the telescopes and their performance \citep{sst1m_performance_paper}, as well as the analysis and reconstruction chain \citep{sst1mpipe_09}. The same dataset was used to measure the gamma-ray angular acceptance, revealing flat sensitivity up to $\sim2.5^\circ$ \citep{2025arXiv250716408J}, which results in a large gamma-ray field of view (FoV). Deep observation of the Cygnus region resulted in spectro-morphological studies of VER J2016+371 and VER J2019+368 \citep{2025arXiv250716408J}. Bottom left panel of Figure~\ref{fig.results} shows the significance map of the VER J2019+368 region. Resolving the two components proved the capabilities of the instrument for the detection of extended sources and morphological studies at multi-TeV energies with a competitive angular resolution of $0.1^\circ$. Observation of a composite supernova remnant CTA 1 was motivated by discrepancies in spectra and morphology of the source measured by different experiments. Analysis of a 30-hour dataset resulted in a source location and spectrum consistent with the LHAASO result, as shown in the upper right panel of Figure~\ref{fig.results}, which may hint an energy-dependent morphology of the source \citep{2025arXiv250926068L}. The blazar monitoring program focused on nearby active galactic nuclei resulted in two Astronomer's Telegrams \#16533 and \#17597, reporting on bright flares of Mrk 421 \citep{2025arXiv250718445M}. Figure~\ref{fig.results}, bottom right, shows the averaged spectrum of Mrk 421 during its increased activity in January 2026.

\section{Prospects for the Galactic center observations from the Southern Hemisphere}

We used the MC model of the telescopes validated with data from the Ond\v{r}ejov Observatory to prepare dedicated Instrument Response Functions (IRFs) for SST-1M located in Malarg\"ue, Argentina, 1420 m a.s.l., with a proposed distance between the telescopes of 120 m. The telescope separation was optimized to provide a trade-off between sensitivity and angular resolution \citep{2025arXiv250716681C}. Using these IRFs, we simulated 250 hours of observations of the Galactic Center and the central molecular region with SST-1M.\footnote{The Galactic Center is visible for 680 dark hour/year above $60^\circ$ zenith angle.} For the source simulations, we used the spectral models of all nearby VHE gamma-ray sources: HESS J1745-290, HESS J1745-303, SNR G0.9+0.1, available in \citet{2024JCAP...10..081A}. The diffuse component was simulated with an exponential cutoff power law (ECPL) spectral model \citep{2020A&A...642A.190M} and with the spatial template derived from a velocity-integrated radio emission map of the region \citep{2016Natur.531..476H}. 

We performed a 3D analysis in \texttt{gammapy} \citep{gammapy} using the \texttt{FoVBackgroundModel} method for background estimation\footnote{\texttt{FoVBackgroundModel} method scales the entire field-of-view background template so that it better matches the observed counts in the source-free regions of the observation.}, considering all bright sources in the region. Sgr A*, SNR G0.9+0.1, and the diffuse emission are significantly detected, while HESS J1745-303 is not detected due to its relatively low flux and angular extension. 

\begin{figure}[t]
\centering
\begin{tabular}{cc}
\includegraphics[width=.48\textwidth]{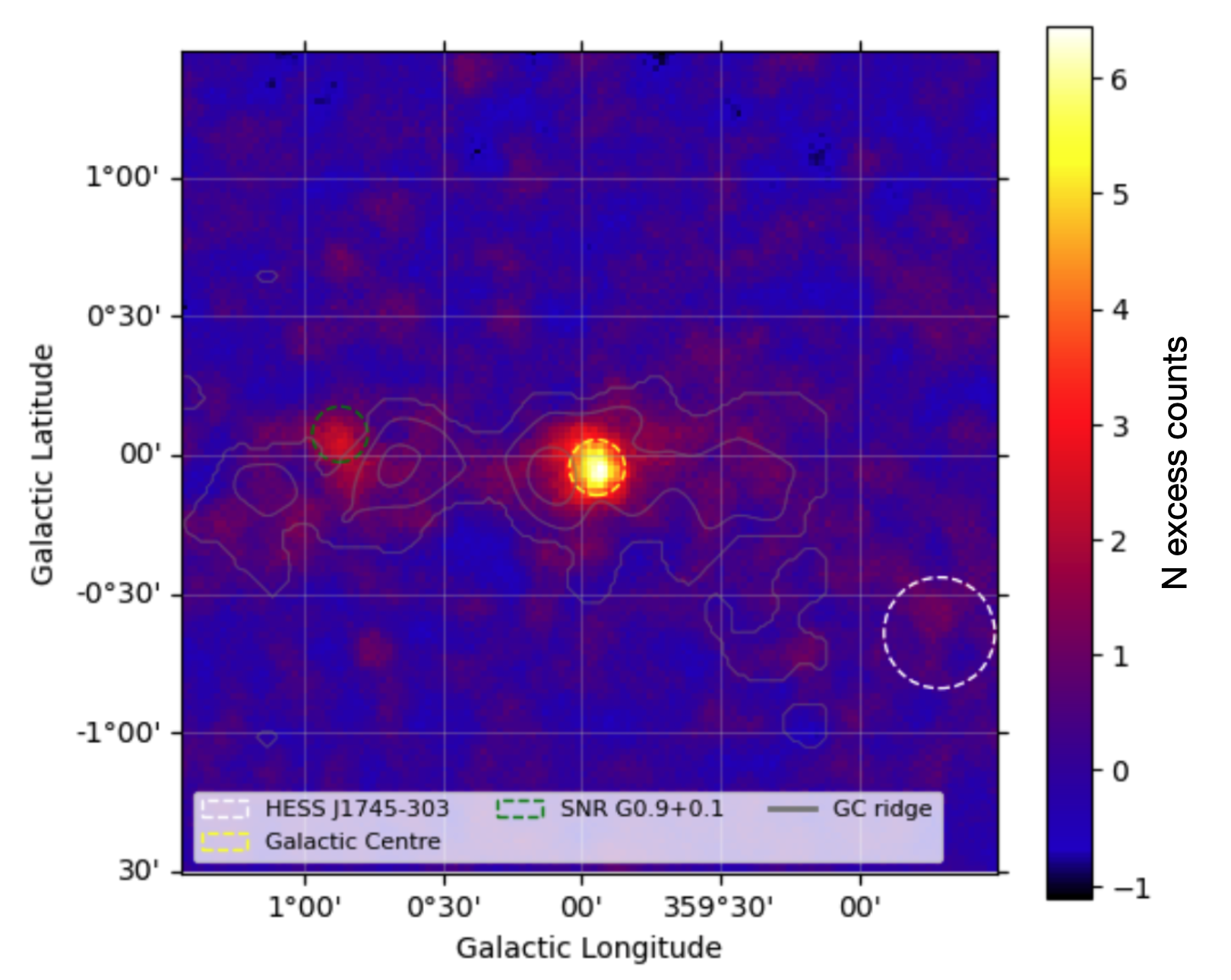} &
\includegraphics[width=.49\textwidth]{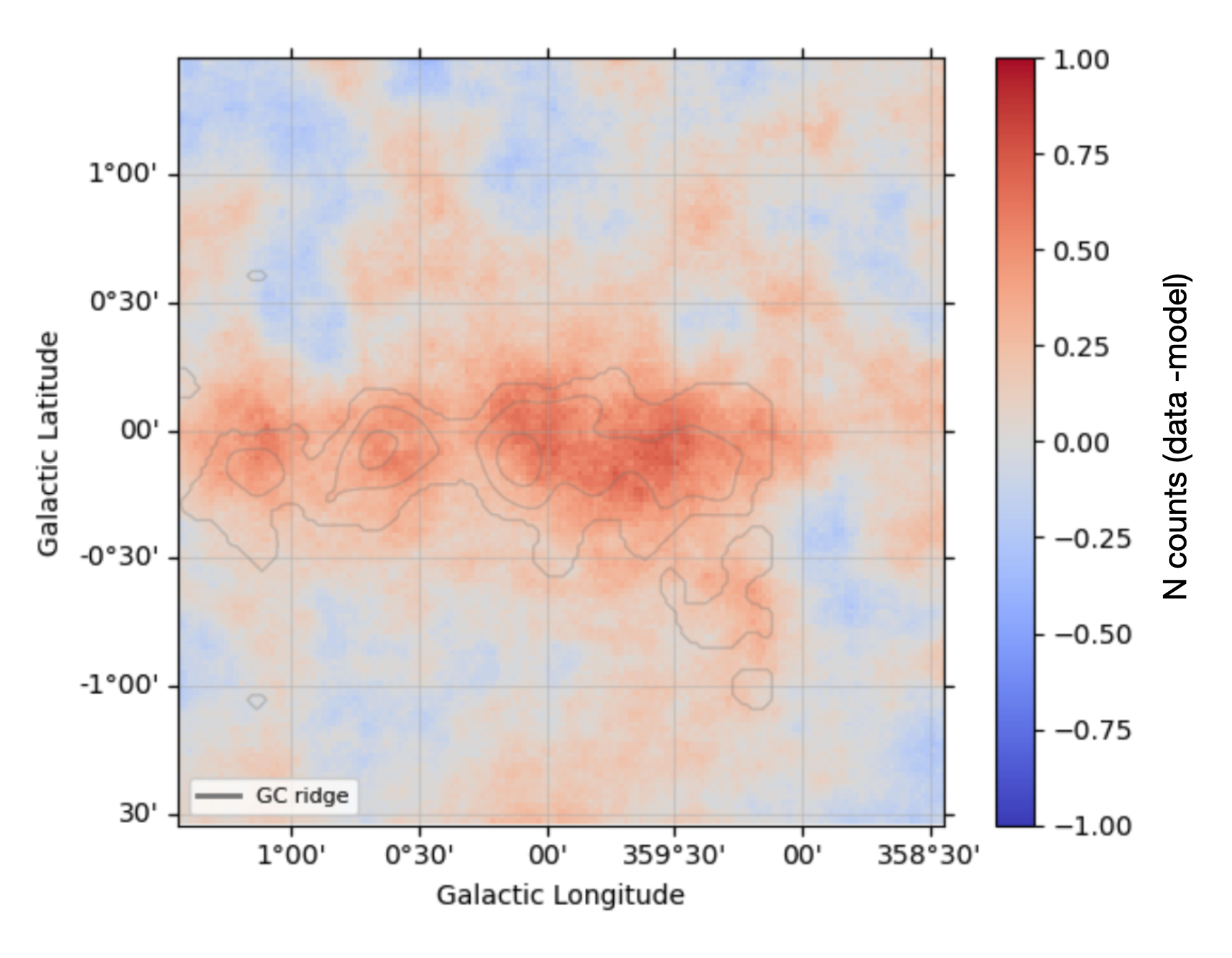} \\
\end{tabular}
\caption{Simulation of 250 hours of SST-1M observation of the Galactic Center region. The simulated spatial model of the diffuse emission is shown as a grey contour. \textit{Left:} The excess map showing the brightest sources in the central molecular region. The color scale represents the number of excess events in each $0.02^\circ$ bin of the map, convoluted with a disk of $0.08^\circ$ radius. \textit{Right:} The residual map after subtraction of all fitted sources (resulting from the 3D modeling of the region) except for the diffuse component emission.}
\label{fig.gcenter_sst1m}
\end{figure}

The map of excess events for 250 hours of observation is shown in the left panel of Fig.~\ref{fig.gcenter_sst1m}. The map is integrated for energies between 1 and 300~TeV, requesting a safety mask on the effective area ($A_\mathrm{eff}> 1\%$ of maximum) and energy bias ($E_\mathrm{bias}< 30\%$) to mitigate systematic uncertainties. Figure~\ref{fig.gcenter_sst1m} (right panel) shows the residual excess map after careful subtraction of all fitted sources in the region except for the diffuse emission component. 

\begin{figure}[t]
\centering
\begin{tabular}{cc}
\includegraphics[width=.49\textwidth]{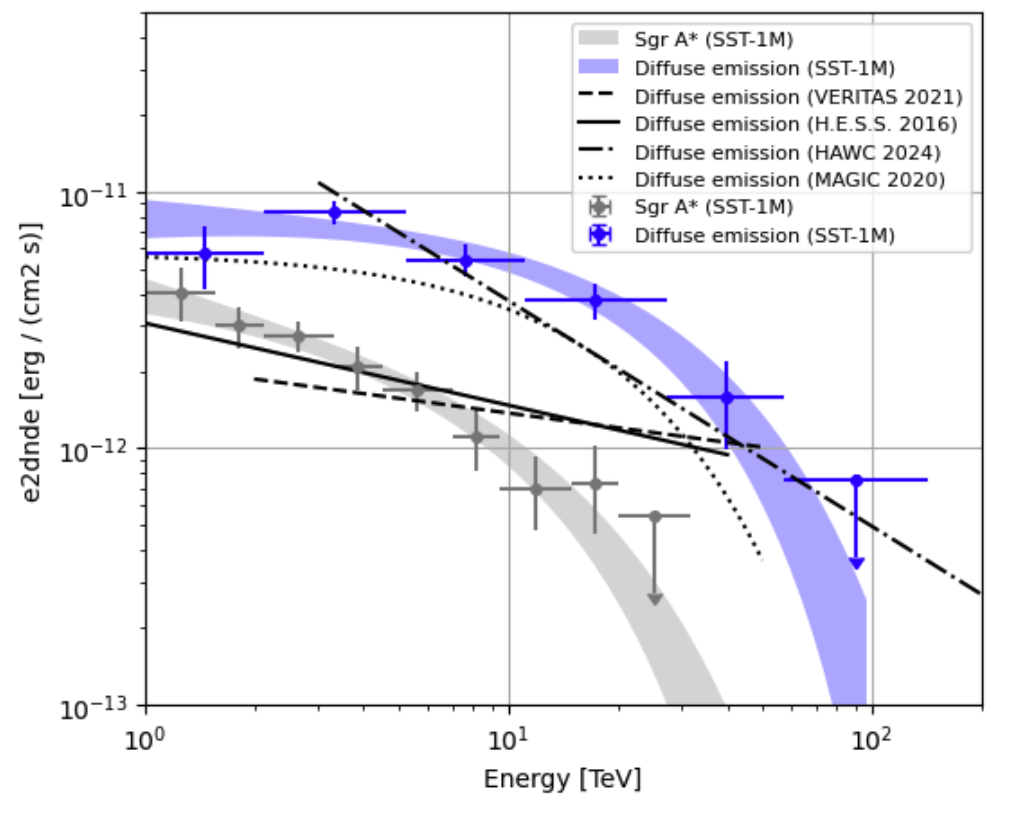} &
\includegraphics[width=.47\textwidth]{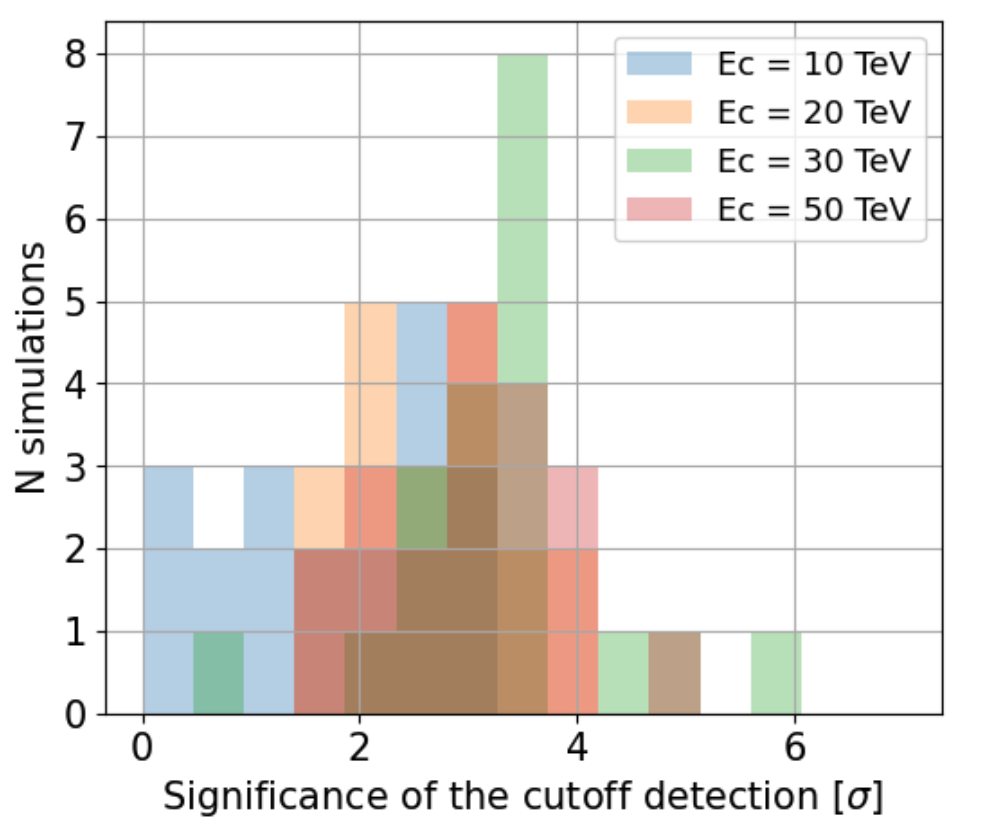} \\
\end{tabular}
\caption{\textit{Left:} Spectrum of the Galactic Center and the diffuse emission as seen with SST-1M in the South. The PL models of the Ridge emission as reported by H.E.S.S. \citep{2016Natur.531..476H}, MAGIC \citep{2020A&A...642A.190M}, VERITAS \citep{2021ApJ...913..115A}, and HAWC \citep{2024ApJ...973L..34A} are also shown for comparison. \textit{Right:} Significance of the cutoff detection for different assumptions of the cutoff energy (20 realisations each).}
\label{fig.sed_cutoff}
\end{figure}

The resulting Spectral Energy Distributions of HESS J1745-290 (Sgr A*) and the diffuse emission as expected to be seen by SST-1M are shown in the left panel of Fig.~\ref{fig.sed_cutoff}, demonstrating capabilities of the SST-1M to explore the spectral shapes of individual sources in the Galactic Center region. In order to evaluate prospects for a cutoff detection, we simulated the diffuse emission with ECPL spectra of different cutoff energies (20 realizations each), and ran a likelihood ratio test assuming PL spectrum (H0) and ECPL spectrum (H1) in the fitted model of the region. The resulting test statistics is given by $\mathrm{TS} = -2\ln({\mathcal{L}(\lambda=0) / \mathcal{L}{(\lambda)})}$, where $\lambda = 1/E_\mathrm{cutoff}$, and $\mathcal{L}{(\lambda)}$ is the maximum likelihood over the full parameter space ($\lambda, \Gamma, \phi_0 $), and restricted parametric space ($\Gamma, \phi_0$), respectively, with $\Gamma$ representing the spectral index and $\phi_0$ the spectral normalization. Using Wilk's theorem \citep{Wilks:1938dza} for nested models with $\Delta\mathrm{dof}=1$ (the single extra parameter being the cutoff energy), $\sqrt{\mathrm{TS}}$ provides the significance of a cutoff detection. The distribution of the cutoff detection significance is shown in the right panel of Fig.~\ref{fig.sed_cutoff}. It shows that SST-1M can see the cutoff with $2-4\sigma$ detection significance, depending on the cutoff energy.

\section{Conclusions}

The SST-1M stereo observatory combines a large FoV and good angular resolution, which makes it an ideal instrument for morphological studies of extended Galactic PeVatron candidates. Early results demonstrate that, even at a preliminary site with challenging atmospheric conditions, the SST-1M is capable of delivering scientifically valuable data. One of the candidate sites considered for the relocation has excellent visibility of the Galactic Center. We performed simulated observations of the region, showing that SST-1M in the proposed location will be able to explore the morphology and spectral shapes of individual sources in the region, including the diffuse emission.

\section{Acknowledgements}\footnotesize

This publication was created as part of the projects funded in Poland by the Minister of Science based on agreements number 2024/WK/03 and DIR/\-WK/2017/12. The construction, calibration, software control and support for operation of the SST-1M cameras is supported by SNF (grants CRSII2\_141877, 20FL21\_154221, CRSII2\_160830, \_166913, 200021-231799), by the Boninchi Foundation and by the Université de Genève, Faculté de Sciences, Département de Physique Nucléaire et Corpusculaire. The Czech partner institutions acknowledge support of the infrastructure and research projects by Ministry of Education, Youth and Sports of the Czech Republic (MEYS) and the European Union funds (EU), MEYS LM2023047, EU/MEYS CZ.02.01.01/00/22\_008/0004632, CZ.02.01.01/00/22\_010/0008598, Co-funded by the European Union (Physics for Future – Grant Agreement No. 101081515), and Czech Science Foundation, GACR 23-05827S.

\bibliographystyle{iaulike}
{\footnotesize
\bibliography{references}
}

\clearpage
\section*{Full Authors List: SST-1M Collaboration}\label{authorlist}
\noindent
C.~Alispach$^1$,
A.~Araudo$^2$,
A.~Bakalov\'a$^2$,
M.~Balbo$^1$,
V.~Beshley$^3$,
J.~Bla\v{z}ek$^2$,
P.~Bo\v{r}il$^2$,
J.~Borkowski$^4$,
T.~Bulik$^6$,
F.~Cadoux$^1$,
S.~Casanova$^5$,
A.~Christov$^2$,
J.~Chudoba$^2$,
L.~Chytka$^7$,
P.~\v{C}echvala$^2$,
P.~D\v{e}dic$^2$,
T.~Dietrich$^{16}$,
Y.~Favre$^1$,
L.~Gantel$^{16}$,
M.~Garczarczyk$^8$,
L.~Gibaud$^9$,
T.~Gieras$^5$,
E.~G{\l}owacki$^9$,
P.~Hamal$^7$,
M.~Heller$^1$,
M.~Hrabovsk\'y$^7$,
P.~Jane\v{c}ek$^2$,
M.~Jel\'inek$^{10}$,
V.~J\'ilek$^7$,
J. Jury\v{s}ek$^{2}$,
V.~Karas$^{11}$,
J.~Kvapil$^{16}$,
B.~Lacave$^1$,
E.~Lyard$^{12}$,
D.~Mand\'at$^2$,
W.~Marek$^5$,
A.~McMullin$^1$,
S.~Michal$^7$,
J.~Micha{\l}owski$^5$,
M.~Miro\'n$^6$,
R.~Moderski$^4$,
T.~Montaruli$^1$,
A.~Muraczewski$^4$,
S.~R.~Muthyala$^2$,
A.~L.~Müller$^2$,
K.~Nalewajski$^5$,
J.~Niemiec$^5$,
M.~Niko{\l}ajuk$^9$,
V.~Novotn\'y$^{2,14}$,
M.~Ostrowski$^{15}$,
M.~Palatka$^2$,
M.~Pech$^2$,
M.~Prouza$^2$,
P.~Schovanek$^2$,
T.~Schultz$^2$
V.~Sliusar$^{12}$,
J.~Srba$^{10}$,
{\L}.~Stawarz$^{15}$,
R.~Sternberger$^8$,
J.~\'{S}wierblewski$^5$,
P.~\'{S}wierk$^5$,
J.~\v{S}trobl$^{10}$,
T.~Tavernier$^2$,
P.~Tr\'avn\'i\v{c}ek$^2$,
A.~Upegui$^{16}$,
M.~Vacula$^7$,
J.~V\'icha$^2$,
R.~Walter$^{12}$,
K.~Zi{\c e}tara$^{15}$
\vspace{5mm}

\noindent
$^1$D\'epartement de Physique Nucl\'eaire, Facult\'e de Sciences, Universit\'e de Gen\`eve, 24 Quai Ernest Ansermet, CH-1205 Gen\`eve, Switzerland.
$^2$FZU - Institute of Physics of the Czech Academy of Sciences, Na Slovance 1999/2, Prague 8, Czech Republic.
$^3$Pidstryhach Institute for Applied Problems of Mechanics and Mathematics, National Academy of Sciences of Ukraine, 3-b Naukova St., 79060, Lviv, Ukraine.
$^4$Nicolaus Copernicus Astronomical Center, Polish Academy of Sciences, ul. Bartycka 18, 00-716 Warsaw, Poland.
$^5$Institute of Nuclear Physics, Polish Academy of Sciences, PL-31342 Krakow, Poland.
$^6$Astronomical Observatory, University of Warsaw, Al. Ujazdowskie 4, 00-478 Warsaw, Poland.
$^7$Palack\'y University Olomouc, Faculty of Science, 17. listopadu 50, Olomouc, Czech Republic.
$^8$Deutsches Elektronen-Synchrotron (DESY) Platanenallee 6, D-15738 Zeuthen, Germany.
$^9$Faculty of Physics, University of Bia{\l}ystok, ul. K. Cio{\l}kowskiego 1L, 15-245 Bia{\l}ystok, Poland.
$^{10}$Astronomical Institute of the Czech Academy of Sciences, Fri\v{c}ova~298, CZ-25165 Ond\v{r}ejov, Czech Republic.
$^{11}$Astronomical Institute of the Czech Academy of Sciences, Bo\v{c}n\'i~II 1401, CZ-14100 Prague, Czech Republic.
$^{12}$D\'epartement d'Astronomie, Facult\'e de Science, Universit\'e de Gen\`eve, Chemin d'Ecogia 16, CH-1290 Versoix, Switzerland.
$^{13}$ETH Zurich, Institute for Particle Physics and Astrophysics, Otto-Stern-Weg 5, 8093 Zurich, Switzerland.
$^{14}$Institute of Particle and Nuclear Physics, Faculty of Mathematics and Physics, Charles University, V Hole\v sovi\v ck\' ach 2, Prague 8, Czech~Republic.
$^{15}$Astronomical Observatory, Jagiellonian University, ul. Orla 171, 30-244 Krakow, Poland.
$^{16}$Haute École du Paysage, d'Ingénierie et d'Architecture de Genève, Haute École Spécialisée de Suisse occidentale, Rue de la Prairie 4, 1202 Geneva, Switzerland.

\end{document}